%% file: main.tex
\documentclass[sigconf]{acmart}

\usepackage{soul}
\usepackage{xspace}
\usepackage{algorithmic}
\usepackage{algorithm}
\usepackage{balance}
\usepackage{enumitem}
\usepackage{tikz}
\usepackage{kotex}
\usepackage{multirow}
\usepackage{subcaption}
\usepackage{booktabs}
\usepackage{hhline}
\usepackage{makecell}
\usepackage[dvipsnames]{xcolor}

\AtBeginDocument{%
  }

\setcopyright{acmlicensed}

\copyrightyear{2026}
\acmYear{2026}
\setcopyright{cc}
\setcctype{by}
\acmConference[SAC '26]{The 41st ACM/SIGAPP Symposium on Applied Computing}{March 23--27, 2026}{Thessaloniki, Greece}
\acmBooktitle{The 41st ACM/SIGAPP Symposium on Applied Computing (SAC '26), March 23--27, 2026, Thessaloniki, Greece}
\acmPrice{}
\acmDOI{10.1145/3748522.3779981}
\acmISBN{979-8-4007-2294-3/2026/03}

\newcommand{\sysname}{\textit{HE-DAP}\xspace}
\newcommand{\ppstat}{\textit{PP-STAT}\xspace}
\newcommand{\hstat}{\textit{HEaaN-STAT}\xspace}

\newcommand{\heaanCpu}{HEaaN-CPU\xspace}
\newcommand{\heaanGpu}{HEaaN-GPU\xspace}
\newcommand{\heaan}{HEaaN\xspace}
\newcommand{\lattigo}{Lattigo\xspace}
\newcommand{\invSqrt}{\textit{CryptoInvSqrt}\xspace}

\begin{document}

\title{\sysname: Homomorphic Encryption-based Dynamic Adaptive Parameter Optimization for Statistical Computation}
\renewcommand{\shorttitle}{\sysname: Homomorphic Encryption-based Dynamic Adaptive Parameter Optimization for Statistical Computation}

\author{Yun-Soo Park}
\email{yunsoo200@inha.edu}
\orcid{0009-0003-4891-2893}
\affiliation{%
  \institution{Inha University}
  \city{Incheon}
  \country{Republic of Korea}
}

\author{Hyunmin Choi}
\email{hyunmin.choi@g.skku.edu}
\orcid{0009-0002-0486-9582}
\affiliation{%
  \institution{NAVER Cloud \& Sungkyunkwan University}
  \city{Seongnam}
  \country{Republic of Korea}
}

\author{Hyoungshick Kim}
\email{hyoung@skku.edu}
\orcid{0000-0002-1605-3866}
\affiliation{%
  \institution{Sungkyunkwan University}
  \city{Suwon}
  \country{Republic of Korea}
}

\author{Mun-Kyu Lee}
\email{mklee@inha.ac.kr}
\orcid{0000-0003-4423-7467}
\affiliation{%
  \institution{Inha University}
  \city{Incheon}
  \country{Republic of Korea}
}

\renewcommand{\shortauthors}{Yun-Soo Park et al.}
\begin{abstract}
Homomorphic encryption (HE) enables privacy-preserving analytics but remains hindered by high computational overhead. We find that the \emph{inverse square root}---a key primitive in many statistical and machine learning workloads---exhibits inconsistent and often suboptimal performance across HE libraries and hardware. This stems from a core trade-off between two costly HE operations: evaluating high-degree Chebyshev polynomials to speed up Newton's method versus performing bootstrapping to manage ciphertext noise. Because their relative costs vary by up to $6\times$ across environments, any fixed configuration proves inherently inefficient.

To address this challenge, we present \sysname, a cross-platform optimization framework that automatically navigates this trade-off. By profiling an environment's unique performance characteristics, \sysname finds the optimal balance between polynomial degree and iteration count to accelerate the encrypted inverse square root computation for a given accuracy target. Our evaluation on \lattigo, \heaanCpu, and \heaanGpu shows that \sysname's adaptive approach yields significant performance gains. It accelerates the core inverse square root computation by up to $2.35\times$ over the fixed configuration in \ppstat while maintaining high numerical accuracy ($\mathrm{MRE}\leq3.1\times10^{-8}$). We further demonstrate that optimizing this fundamental building block directly enhances the end-to-end performance of complex statistical analyses, confirming the practical benefits of our environment-aware approach. By automatically adapting to heterogeneous execution environments, \sysname demonstrates that principled parameter optimization can make privacy-preserving statistical analytics practical at scale.
\end{abstract}

\ccsdesc[500]{Security and privacy~Privacy-preserving protocols}


\keywords{Homomorphic encryption, Parameter optimization, Privacy}

\maketitle

\begingroup
\makeatletter
\renewcommand\@makefnmark{}
\makeatother
\footnotetext{This paper was presented at \textit{the 41st ACM/SIGAPP Symposium On Applied Computing (SAC'26)}.}
\endgroup
  
\input{section/1_Introduction}
\input{section/2_Background}
\input{section/4_Limits_of_ppstat}
\input{section/5_Proposed_method}

\input{section/6_Experiments}

\input{section/7_Discussion}
\input{section/3_Related_work}
\input{section/Security_Analysis}
\input{section/8_Conclusion}
\input{section/Acknowledgments}
\bibliographystyle{ACM-Reference-Format}
\balance
\bibliography{sample-base}
\appendix

\end{document}

%% file: section/1_Introduction.tex
\section{Introduction}
\label{sec:introduction}
In modern cloud environments, outsourcing data processing to remote servers raises concerns about sensitive information leakage. To address these, homomorphic encryption (HE)~\cite{gentry2009fully,bgv,tfhe,cheon2017homomorphic}, which enables computation on encrypted data, has emerged as an effective solution for privacy-preserving cloud services~\cite{ibm2024fhe,microsoft2021passwordmonitor,apple2023he,redhat2024fhe}.

In this paper, we consider an HE-based advanced statistical analysis service that enables secure computation on encrypted client data performed by the server. A key technical challenge in such analysis is the efficient computation of the inverse square root over encrypted data. Several studies have investigated HE-based constructions for this operation. Panda~\cite{panda2022polynomial} proposed the Pivot–Tangent method, which offers low multiplicative depth at the expense of limited precision. In contrast, Lee et al.~\cite{lee2023heaan} introduced an approach based on Newton’s method, known for its high precision but typically requiring greater multiplicative depth. 
PP-STAT~\cite{choi2025pp} also employed Newton’s method and further improved its convergence by using the Chebyshev approximation to determine a better initial value. With this improved inverse square root algorithm, PP-STAT achieved significant acceleration in five core statistical functions---Z-score normalization, skewness, kurtosis, coefficient of variation, and the Pearson correlation coefficient.

Optimizing the inverse square root operation in HE requires resolving a fundamental trade-off between the number of bootstraps for noise management and the degree of the approximation polynomial, which dictates the required multiplicative depth. The optimal balance for this trade-off is highly contingent upon the specific HE library and hardware in use. However, despite this high environmental dependency, prior work~\cite{lee2023heaan, panda2022polynomial, qu2023improvements, choi2025pp} has been limited to proof-of-concept evaluations within a single, static environment. Consequently, existing frameworks such as HEaaN-STAT~\cite{lee2023heaan} and PP-STAT~\cite{choi2025pp} rely on static parameters tailored to a specific library, hindering their generalizability. For instance, PP-STAT hard-codes a specific polynomial degree and a fixed number of Newton's iterations, which are optimized for the Lattigo library~\cite{lattigo}. Our analysis reveals, in stark contrast, that a completely different set of values yields optimal performance for the HEaaN library~\cite{heaan} in both CPU and GPU environments.

To overcome this limitation, we propose \sysname, an efficient \textbf{\underline{H}}omomorphic \textbf{\underline{E}}ncryption-based \textbf{\underline{D}}ynamic \textbf{\underline{A}}daptive \textbf{\underline{P}}arameter optimizer that automatically determines the ideal parameters for a given HE library and hardware configuration. \sysname runs once during the installation of a target framework (e.g. PP-STAT) and identifies the parameter configuration that minimizes latency while preserving the target precision for the inverse square root computation. With the integration of \sysname, the runtime of inverse square root operations in the \heaanGpu library is reduced by up to 2.35$\times$ compared to the framework’s original static configuration, and experiments on statistical measures across three distinct settings (\lattigo, \heaanCpu, and \heaanGpu) confirm that \sysname consistently discovers the optimal parameter configuration for both CPU and GPU environments.

{Our main contributions are as follows: }
\begin{itemize} 
\item We introduce \sysname, a novel adaptive optimizer that automatically configures the inverse square root operation in HE, for optimal performance across diverse HE libraries and hardware environments.

\item We demonstrate that \sysname with \ppstat in the \heaanGpu achieves a significant performance gain, with up to a 2.35 times speedup for \ppstat over static configurations, while maintaining numerical accuracy.

\item We provide a comprehensive evaluation of \sysname on four core statistical measures across three distinct HE settings (\lattigo, \heaanCpu, and \heaanGpu), demonstrating its practical effectiveness and robustness.

\item We release the full implementation of \sysname as open-source to ensure reproducibility. The code is available at \url{https://github.com/hm-choi/he_dap}.
\end{itemize}

%% file: section/2_Background.tex
\section{Background}
\label{sec:background}

\subsection{Homomorphic encryption}
\label{subsec:he_background}
Homomorphic encryption (HE) enables computation on encrypted data. We focus on Fully Homomorphic Encryption (FHE), which supports an arbitrary number of operations. We use the Cheon--Kim--Kim--Song (CKKS) scheme \cite{cheon2017homomorphic}, which performs approximate arithmetic on vectors of real or complex numbers in a SIMD manner.

We denote element-wise operations as follows: addition ($Add$), subtraction ($Sub$), and multiplication ($Mul$). The subscripts $_P$ and $_C$ distinguish between operations involving a plaintext ($P$) and a ciphertext ($C$), such as $Add_P(C, P)$, and those between two ciphertexts, like $Add_C(C_1, C_2)$.
Multiplications ($Mul_P$, $Mul_C$) consume a ciphertext's \textit{level}. A ciphertext's level represents the number of further multiplications it can undergo before becoming indecipherable. This limitation is addressed by bootstrapping ($BTS$)~\cite{cheon2018bts}, a procedure that resets a ciphertext's level to a specific value, $l_{\text{after}BTS}$, effectively enabling a virtually unlimited number of multiplications. However, $BTS$ is computationally expensive compared to basic operations and can only be performed if the ciphertext's level is at least $l_{BTS}$. Consequently, a primary challenge in designing HE-based algorithms is to minimize the number of $BTS$ operations~\cite{paindavoine2015minbts, cheon2024dacapo, liu2025resbm, cheon2025halo}.

\noindent\textbf{Lattigo.}
Lattigo~\cite{lattigo} is a Go-based open-source HE library supporting BFV~\cite{fv}, BGV~\cite{bgv} and CKKS on standard CPU.
Its CKKS implementation provides flexible parameter configuration. In our experiments, we use a ring dimension of $N=2^{16}$, maximum depth 27, scaling factor $\Delta=2^{50}$, and total modulus size $\log_2(PQ)=1553$, which ensure a 128-bit security level under the standard RLWE 
assumptions~\cite{bossuat2024security}. Under this configuration, $l_{BTS}$ and $l_{\text{after}BTS}$ are set to 0 and 11, respectively, and the maximum level $l_{\max}$ is also 11.

\noindent\textbf{HEaaN.}
HEaaN~\cite{heaan} is an HE library supporting CKKS. It provides two variants, \heaanCpu and \heaanGpu, optimized for CPU and GPU environments, respectively. 
\footnote{In this paper, we utilize the \textit{heaan-stat:1.0.0-gpu} Docker image, provided by CryptoLab and publicly available at ~\cite{heaanDocker}, for \heaanCpu and \heaanGpu.}
Among various CKKS parameter configurations, we focus on the \textit{Full Grande b} (FGb) setting, which supports bootstrapping. This configuration uses $N=2^{16}$, maximum depth 24, $\Delta=2^{42}$, and $\log_2(PQ)=1555$, ensuring 128-bit security under the standard RLWE assumptions~\cite{bossuat2024security}. The parameters $l_{\max}$, $l_{BTS}$, and $l_{\text{after}BTS}$ are set to 12, 3, and 12, respectively.

\subsection{Newton–Raphson method}
\label{subsec:newton_method}
The Newton–Raphson method~\cite{akram2015newton} (a.k.a. Newton’s method) is a root-finding algorithm that produces successively better approximations to the roots of a real-valued function $f$. To compute a specific value $b$ via Newton's method, we first define a function $f$ such that $b$ is a root of $f$, and then iteratively compute
$y_{n+1} = y_n - f(y_n)/f'(y_n)$
until the error between $y_{n+1}$ and $b$ becomes sufficiently small. In this paper, we compute the inverse square root $b = 1/\sqrt{a}$ with $a > 0$ using Newton’s method applied to $f(x) = x^{-2} - a$, where $f(b) = 0$. The procedure, presented in Algorithm 1 in \cite{choi2025pp}, iteratively computes
$y_{n+1} = 0.5 y_{n}  (3 - a y_{n}^{2})$
for $n \geq 0$, with the initial value $y_0$. Newton's method converges rapidly when $y_0$ is close to $b$, making the selection of a good initial value critical for fast convergence.

\subsection{\invSqrt}
\label{subsec:crypto_invsqrt}
PP-STAT~\cite{choi2025pp} introduced \invSqrt,
the state-of-the-art method for computing the inverse square root over the CKKS scheme. (See Algorithm~\ref{alg:invSqrt} below.)
It employs a Chebyshev approximation~\cite{hearnandez2001cheb} to determine a good initial value $y_{0}$ for Newton’s method. 
Algorithm~\ref{alg:invSqrt} operates correctly for inputs within the domain $[0, 1]$;
however, this assumption is overly restrictive in practical scenarios.
To relax this constraint, we adopt the pre-normalization scaling method introduced in Section~6.1 in \cite{choi2025pp}.
Throughout this paper, we denote the scaling constant as $B$.

\begin{algorithm}[!ht]
\small
\caption{Inverse Square Root Computation (\invSqrt)}
\label{alg:invSqrt}
\begin{algorithmic}[1]
\STATE {\bfseries Input:} 
$ct$ (ciphertext for $a$ (Assume $a \in [0, 1]$), $d$ (degree of Chebyshev approximation),
$n$ (number of Newton's iterations)
\STATE {\bfseries Output:}
$ct_{n}$ (ciphertext of approximated $1/\sqrt{a}$)
\STATE $ct_{0} \leftarrow \text{ChebyshevApprox}(ct, d)$
\STATE $ct_{0} \leftarrow \texttt{BTS}(ct_{0})$
\STATE $ct_{x} \leftarrow Mul_P(ct, 0.5)$
\FOR{$i=1$ {\bfseries to} $n$}
    \STATE $tmp_{\alpha} \leftarrow Mul_P(ct_{i-1}, 1.5)$
    \STATE $tmp_{\beta} \leftarrow Mul_C(ct_{x}, ct_{i-1})$
    \STATE $ct_{i-1} \leftarrow Mul_C(ct_{i-1}, ct_{i-1})$
    \STATE $tmp_{\beta} \leftarrow Mul_C(tmp_{\beta}, ct_{i-1})$
    \STATE $ct_{i} \leftarrow Sub_C(tmp_{\alpha}, tmp_{\beta})$
\ENDFOR
\STATE \textbf{return} $ct_{n}$
\end{algorithmic}
\end{algorithm}

%% file: section/4_Limits_of_ppstat.tex
\section{Necessity of Parameter Tuning}
\label{sec:limitsofppstat}

\invSqrt in PP-STAT employs a fixed Chebyshev polynomial degree of $2^9-2$ and a fixed number 6 of Newton iterations. However, these parameters, not being derived from comprehensive experimentation, are not universally optimal across environments.

A key limitation of PP-STAT is its assumption that the input ciphertext level $l_{input}$ is $l_{max}$, which often does not hold in practice. Consequently, a preliminary bootstrapping, denoted as Pre-BTS, becomes necessary whenever $l_{input}$ is insufficient to accommodate the depth of the Chebyshev polynomial evaluation.

Furthermore, the empirical validation in PP-STAT was confined to a CPU environment using \lattigo. In \lattigo, bootstrapping is significantly more time-consuming than Chebyshev polynomial evaluation. Consequently, using a high-degree polynomial to secure a precise initial value is an effective strategy, as it minimizes the number of costly Newton iterations. In contrast, the performance trade-off is inverted in GPU-accelerated libraries like \heaanGpu, where bootstrapping is remarkably fast and often outperforms the evaluation of high-degree Chebyshev polynomials. This shift renders the strategy of minimizing iterations at the cost of a complex initial approximation less effective.

Table~\ref{table:cheb_comparison} quantifies this performance difference in \heaanGpu. 
It compares the runtimes for Chebyshev polynomial evaluation (\texttt{Cheb}) against $BTS$. The data reveal a crossover point: for degrees $\geq 2^6-2$, the Chebyshev evaluation becomes slower than $BTS$.
At the degree of $2^9-2$ used by PP-STAT, it is nearly six times slower.

\begin{table}[!t]
\caption{Runtimes of Chebyshev polynomial evaluation (\texttt{Cheb}) and $BTS$ in the FGb setting (average of $50$ trials, standard deviations in parentheses). $BTS$ runtime $(a)$ is $81.13$ms.}
\centering
\footnotesize
\renewcommand{\arraystretch}{1.2}
\setlength{\tabcolsep}{5pt}
\begin{tabular}{ccccccc}
\toprule
\textbf{Degree} & $2^4-2$ & $2^5-2$ & $2^6-2$ & $2^7-2$ & $2^8-2$ & $2^9-2$ \\
\midrule
$(b)$ \textbf{\texttt{Cheb}} (ms) 
& \makecell{14.77 \\ (0.96)} 
& \makecell{45.61 \\ (0.32)} 
& \makecell{84.62 \\ (0.58)} 
& \makecell{152.17 \\ (0.63)} 
& \makecell{277.73 \\ (0.39)} 
& \makecell{480.91 \\ (0.97)} \\

\textbf{Ratio: $(b)/(a)$} 
& 0.18$\times$ 
& 0.56$\times$ 
& 1.04$\times$ 
& 1.88$\times$ 
& 3.42$\times$ 
& 5.93$\times$ \\
\bottomrule
\end{tabular}
\label{table:cheb_comparison}
\end{table}

Another critical constraint is circuit depth. Evaluating a Chebyshev polynomial of degree $2^9-2$ requires a multiplicative depth of 9, necessitating an input ciphertext with at least 11 levels remaining. This is because several statistical operations, such as Z-score normalization, kurtosis, and the Pearson correlation coefficient, require the square root of the variance term in their denominators. Consequently, \invSqrt cannot be directly applied to ciphertexts from preceding analyses if their remaining level is below this threshold. The workaround---adding a bootstrapping operation before \invSqrt---introduces significant performance overhead. In such cases, using a lower-degree Chebyshev polynomial can be more efficient by avoiding this costly Pre-BTS step.

In conclusion, the fixed-parameter approach of PP-STAT is not optimal. For efficient execution of \invSqrt, the degree of the Chebyshev polynomial and the number of Newton iterations must be co-optimized. This optimization should holistically consider the execution environment (e.g., CPU vs. GPU), the specific HE library, and the available level of the input ciphertext.

%% file: section/5_Proposed_method.tex
\section{Proposed Method}
\label{sec:proposedmethod}

In this section, we propose \sysname, a new method that determines the optimal parameters of \textit{CryptoInvSqrt} on the target platform.

\subsection{Overview of \sysname}
\label{subsec:overview}

\begin{figure}[!h]
\centerline{\includegraphics[width=1.0\columnwidth]{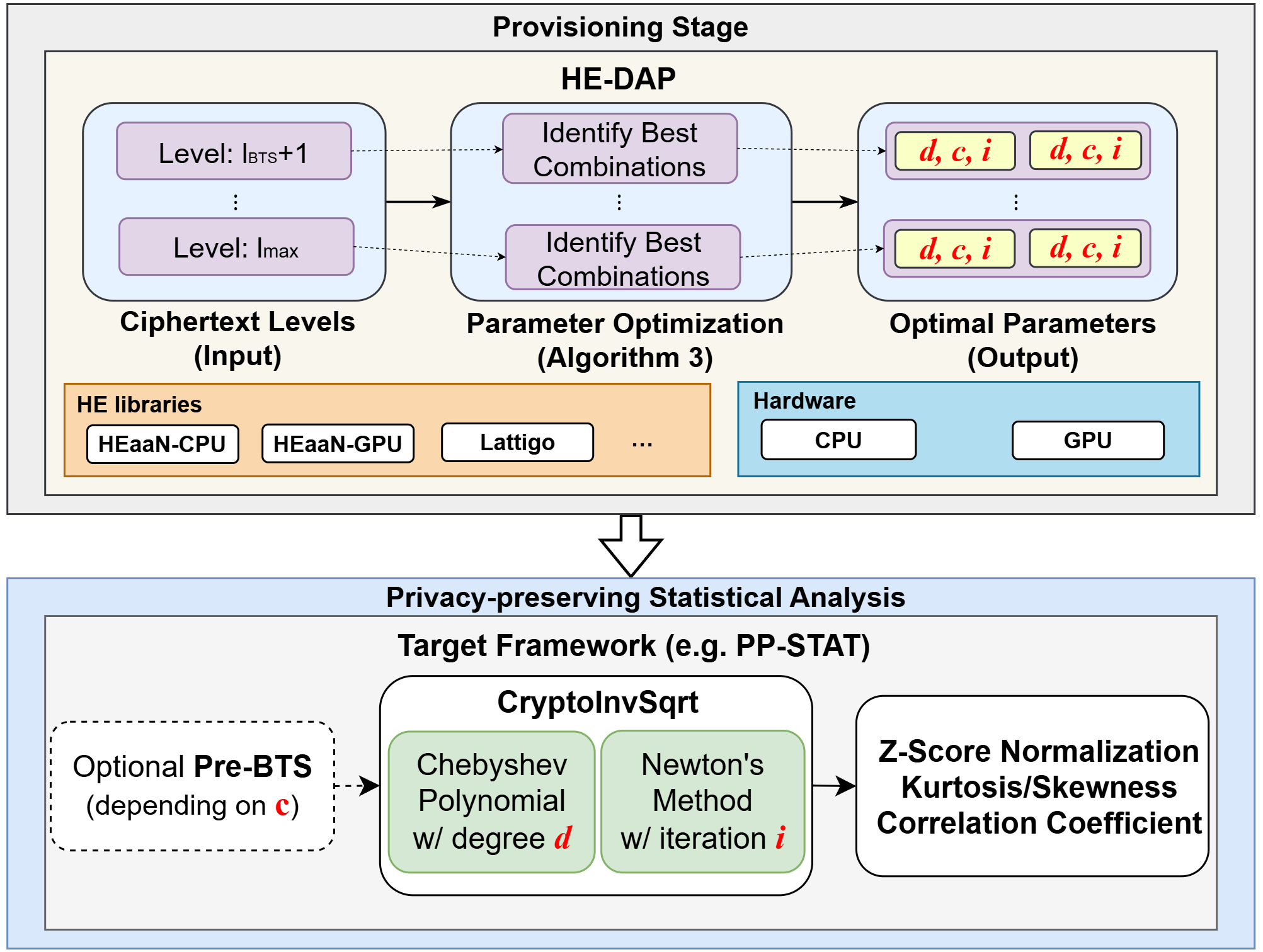}}
\caption{Overview of the privacy-preserving statistical analysis using \sysname.}
\label{fig:overview}
\end{figure}

Figure~\ref{fig:overview} presents an overview of our proposed optimization method, \sysname, for \textit{CryptoInvSqrt}. As shown, \sysname performs a provisioning step to tune the optimal parameters of \textit{CryptoInvSqrt} on the target server framework before execution.

First, \sysname profiles all possible input ciphertext levels, ranging from $l_{BTS}+1$ to $l_{\max}$. Next, for each profiled ciphertext level, it evaluates all possible combinations of the Chebyshev polynomial degree $d$, the number of Newton’s method iterations $i$, and the Pre-BTS indicator $c$, where $c=1$ denotes that $BTS$ is applied to the input ciphertext before executing \textit{CryptoInvSqrt}, and $c=0$ denotes that it is not. When the input ciphertext level is too low, it may cause excessive invocation of $BTS$ during Newton's method, resulting in inefficiency. In such a case, performing $BTS$ in advance can be more advantageous, which leads us to consider Pre-BTS as an optimization option.

Finally, \sysname determines the optimal number of iterations $i$ for each combination of $(d,c)$, defined as the smallest iteration count for which the mean relative error (MRE) is within an allowable range. 
For each level $l$, two $(d,c,i)$ combinations are recorded: the first minimizes the runtime 
under the above constraint on MRE,
and the second minimizes the runtime regardless of the MRE. The former is chosen when accuracy is prioritized, whereas the latter is chosen when speed is prioritized.

For an actual statistical analysis task, a client encrypts its data using HE and sends the ciphertexts 
to the server. The server then performs the statistical computation on the ciphertext domain using the identified optimal parameters, as illustrated in the lower part of Figure ~\ref{fig:overview}, and returns the resulting ciphertext to the client. Finally, the client decrypts it to obtain the final value.
\subsection{Parameter Selection for \invSqrt}
\label{subsec:optimal_params}

In this section, we explain how to find the optimal parameters of \invSqrt. All algorithms described in this section are executed by the server during the provisioning stage. Through this process, the server estimates the expected MRE and runtime during the execution of \invSqrt.

Algorithm ~\ref{alg:optimizing_iteration} describes \textsc{GetOptIter}, a subroutine used in the main identifier algorithm (Algorithm~\ref{alg:optimizing_params}). 
It determines the optimal number of iterations for a server-generated dummy ciphertext with a specific preset level.

\begin{algorithm}[!h]
\small
\caption{\textsc{GetOptIter}}
\label{alg:optimizing_iteration}
\begin{algorithmic}[1]
\STATE {\bfseries Input:} 
        $ct$ (input ciphertext),
        $ans$ (array of true inverse square root values),
        $d$ (degree of Chebyshev polynomial),
        $i_{\max}$ (maximum number of iterations),
        $\delta$ (hyperparameter for threshold decision)
\STATE {\bfseries Output:} 
        $I$ (optimal number of iterations),
        $m$ (MRE at iteration $I$),
        $t$ (runtime at iteration $I$)

\STATE $\mathbf{M}, \mathbf{T} \in \mathbb{R}^{i_{\max}}$
\STATE \textit{Start} $\gets$ \textsc{Current Time}
\STATE $ct_{y_0} \gets \textsc{Cheb}(ct, d)$
\FOR{$i=1$ {\bfseries to} $i_{\max}$}
    \STATE $ct_{y_{i}} \gets$ \textsc{NTIter}$(ct_{y_{i-1}})$
    \STATE $\mathbf{M}[i] \gets$ \textsc{GetMRE}(Dec($ct_{y_{i}}$), $ans$)
    \STATE $\mathbf{T}[i] \gets$ \textsc{Since}(\textit{Start})
\ENDFOR

\STATE $\mathrm{MRE}_{\min} \gets \min(\mathbf{M})$
\STATE $\ell \gets \lfloor \log_{10}(\mathrm{MRE}_{\min}) \rfloor$
\STATE $\alpha \gets \mathrm{MRE_{min}} / 10^{\ell} $
\STATE $\mathrm{MRE_{\delta}} \gets (\lfloor \alpha \rfloor + \delta)\cdot 10^{\ell}$
\STATE $I \gets \min \{\, i \;|\; \mathbf{M}[i] \leq \mathrm{MRE_{\delta}} \,\}$
\STATE \textbf{return} $I, \mathbf{M}[I], \mathbf{T}[I]$
\end{algorithmic}
\end{algorithm}

First, $\mathbf{M}$ and $\mathbf{T}$ are arrays that will store the estimated MRE and runtime for each Newton iteration, respectively. In line~4, the current time is stored in \textit{Start}, after which 
the function \textsc{Cheb} approximates the inverse square root of ciphertext $ct$ with a degree-$d$ Chebyshev polynomial.
The result $ct_{y_0}$ is then used as the initial value for Newton's method, as in \textit{CryptoInvSqrt}.

Next, Newton's iterations are invoked up to $i_{\max}$ times. For each iteration, the function \textsc{NTIter} is executed, performing a single Newton iteration to compute the inverse square root in CKKS. The MRE and accumulated runtime at each iteration are then measured using the \textsc{GetMRE} and \textsc{Since} functions, and are stored in the arrays $\mathbf{M}$ and $\mathbf{T}$, respectively.
We remark that to decrypt $ct_{y_i}$ by Dec($ct_{y_{i}}$), a secret key is required.
For the provisioning stage, the server uses its own public and secret keys, which are different from the client's keys used for the actual statistical analysis. The input $ct$ for Algorithm~\ref{alg:optimizing_iteration} is a dummy ciphertext generated with the server's public key, and $ct_{y_i}$ is decrypted with the corresponding secret key.

After all iterations are completed, the algorithm determines the optimal number of iterations. It first identifies the minimum MRE among all recorded values, denoted by $\mathrm{MRE_{\min}}$. This value can be expressed as $\mathrm{MRE_{\min}} = \alpha \cdot 10^{\ell}$, where $\ell$ is the exponent and $\alpha$ is the mantissa. The threshold MRE, $\mathrm{MRE_{\delta}}$, is then defined based on $\alpha$ and $\ell$, with $\delta$ serving as the hyperparameter. Finally, the smallest iteration number $I$ such that $\mathbf{M}[I] \leq \mathrm{MRE_{\delta}}$ is returned along with the corresponding MRE $\mathbf{M}[I]$ and runtime $\mathbf{T}[I]$.

Algorithm ~\ref{alg:optimizing_params} identifies the optimal parameter combinations $(d,c,i)$ for each ciphertext level $l$, where $d$ is the degree of the Chebyshev polynomial, $c$ is the Pre-BTS indicator, and $i$ is the number of iterations. Leveraging the \textsc{GetOptIter} function, the optimal $i$ is determined for $d$ and $c$ at ciphertext level $l$. The resulting tuple $(d,c,i,m,t)$ is then stored in the map $D$, where $m$ denotes the MRE, and $t$ denotes the execution time. In this map, each key $l$ is mapped to a set of tuples $(d,c,i,m,t)$. 
As explained in Section~\ref {subsec:overview}, we define the range of all possible input ciphertext levels as $[l_{BTS}+1, l_{\max}]$. The lower bound is increased by $1$ from $l_{BTS}$ to ensure that scaling can be applied to the input ciphertext.

\begin{algorithm}[!t]
\small
\caption{\textit{Optimizing the parameters for \textit{CryptoInvSqrt}}}
\label{alg:optimizing_params}
\begin{algorithmic}[1]
\STATE {\bfseries Input:} 
        $d_{\min}$, $d_{\max}$ (range parameters for Chebyshev polynomial degrees)
        $i_{\max}$ (maximum number of iterations),
        $A$, $B$ (minimum and maximum values of input range),
        $N$ (ring dimension of CKKS),
        $\theta$ (hyperparameter for threshold decision),
        $\delta$ (threshold hyperparameter for \textsc{GetOptIter})
        
\STATE {\bfseries Output:} 
        $R$ (two-dimensional array of optimal parameter tuples for each ciphertext level)

\STATE $ans \gets$ Array of inverse square roots of $\textsc{linspace}(A,B,N/2)$
\STATE $D \gets$ map<key: $l$, value: set of tuples $(d,c,i,m,t)$ >
\STATE $R \gets$ map<key: $l$, value:$\big[(d_0,c_0,i_0),\; (d_1,i_1,c_1)\big]$>

\FOR{$l=l_{BTS}+1$ {\bfseries to} $l_{\max}$}
    \FOR{$d_e=d_{\min}$ {\bfseries to} $d_{\max}$}
        \STATE $ct \gets Enc(\textsc{linspace}(A,B,N/2), l)$
        \IF{$l \geq l_{BTS}+3$}
            \STATE $i,m,t \gets \textsc{GetOptIter}(ct, ans, 2^{d_e}-2, i_{\max}, \delta)$
            \STATE Insert $(2^{d_e}-2,0,i,m,t)$ into $D[l]$
        \ENDIF
        \IF{$l < l_{\text{after}BTS}-1$}
            \STATE $ct \gets BTS(ct)$
            \STATE $i,m,t \gets \textsc{GetOptIter}(ct, ans, 2^{d_e}-2, i_{\max}, \delta)$
            \STATE Insert $(2^{d_e}-2,1,i,m,t)$ into $D[l]$
        \ENDIF
    \ENDFOR
    \STATE $m_{\min} \gets \min(\{m|(d,c,i,m,t) \in D[l]\})$
    \STATE $\ell \gets \lfloor \log_{10}(m_{\min}) \rfloor$
    \STATE $\alpha \gets m_{\min} / 10^{\ell} $
    \STATE $m_\theta \gets (\lfloor \alpha \rfloor + \theta)\cdot 10^{\ell}$
    \STATE $u_1 \gets {\rm argmin}_{(d,c,i,m,t) \in D[l] , m \leq m_\theta } (t)$
    \STATE $u_1 \gets (d,c,i)$ from $u_1$
    \STATE $u_2 \gets {\rm argmin}_{(d,c,i,m,t) \in D[l]} (t)$\\
    \STATE $u_2 \gets (d,c,i)$ from $u_2$
    \STATE $R[l] \gets [u_1, u_2]$

\ENDFOR

\STATE \textbf{return} $R$
\end{algorithmic}
\end{algorithm}

In line~3,
$\textsc{linspace}(A,B,N/2)$ is an array of $N/2$ evenly spaced points between $A$ and $B$.
Line 3 computes the
element-wise inverse square roots for $N/2$ elements
and the result is assigned to array $ans$. 
In line~8, we generate a level-$l$ ciphertext, encrypting $N/2$ elements of $\textsc{linspace}(A,B,N/2)$. 
Lines~9 and~13 specify the conditions under which $c=0$ and $c=1$ are evaluated, respectively.
Line ~9 checks whether the input ciphertext level remains greater than or equal to $l_{BTS}$ after consuming one level for scaling and two levels for a single Newton iteration. This condition must hold to continue the Newton iterations when $BTS$ is not performed prior to \invSqrt, which corresponds to $c=0$.
Line ~13, in contrast, corresponds to the condition for $c=1$. Since the allowed input range of $BTS$ is $[-1,1]$, the ciphertext must be scaled into this range for performing BTS. Then, BTS is applied, and the ciphertext is scaled back to its original range. For Pre-BTS to be advantageous, the level $l_{\text{after}BTS}-1$ must exceed the level before $BTS$. Line ~13 confirms this condition, thereby ensuring that Pre-BTS is applied only when it is advantageous.
We remark that all $l$'s of interest fall into at least one case in either line 9 or 13. For some $l$, both cases may be relevant.

To determine the optimal parameters of \invSqrt at level $l$, the algorithm first identifies the minimum MRE among all tuples stored in $D[l]$, denoted as $m_{\min}$. Lines 20--22 computes
the threshold MRE $m_{\theta}$ using the exponent and mantissa of $m_{\min}$ with $\theta$ as the hyperparameter. 
The optimal parameters $(d_0,i_0,c_0,m_0,t_0)$ are determined as the tuple $(d,i,c,m,t)$ in $D[l]$ with the smallest $t$ among those satisfying $m \leq m_{\theta}$. 
Furthermore, the optimal parameters $(d_1,i_1,c_1,m_1,t_1)$ are determined as the tuple $(d,i,c,m,t)$ in $D[l]$ with only the smallest $t$, independent of any constraint on $m$.

In summary, \sysname identifies the optimal parameter combinations for each possible input ciphertext level through Algorithm ~\ref{alg:optimizing_iteration} and ~\ref{alg:optimizing_params}. Depending on whether the user prioritizes accuracy or efficiency, \textit{CyprtoInvSqrt} leverages $(d_0,c_0,i_0,m_0,t_0)$ to minimize the runtime
under a bounded-MRE constraint, whereas it leverages $(d_1,c_1,i_1,m_1,t_1)$ to minimize the runtime regardless of the MRE.

\subsection{Analysis of identified optimal parameters}
\label{subsec:analysis_optimal_params}
In this section, we analyze the optimal parameters for \textit{CryptoInvSqrt} found by the method described in Section ~\ref{subsec:optimal_params}. In PP-STAT~\cite{choi2025pp}, with \lattigo as the underlying library, $d$ was fixed at $2^9-2$. However, as noted in Table~\ref {table:cheb_comparison}, evaluating a polynomial of this degree in \heaanGpu is about $6\times$ slower than $BTS$. Therefore, in the provisioning stage, degrees smaller than $2^9 - 2$ must also be considered, and we set the search range of the degree to $2^{d_e}-2$, where $d_e \in\{4,5,6,7,8,9\}$. When the evaluation result with $d_e=4$ is used as the initial value, at least 12 iterations are required to converge in the plaintext setting. Hence, we set $i_{max}=15$.

Due to space limitations, we provide the full analysis only for \lattigo. We use the parameter settings described in Section~\ref{subsec:he_background}. The input range of the inverse square root is set to $[0.001, 100]$, identical to that used in PP-STAT.

\begin{figure*}[t]
\centering
  \begin{subfigure}{0.49\linewidth}
    \centering
    \includegraphics[width=0.95\linewidth]{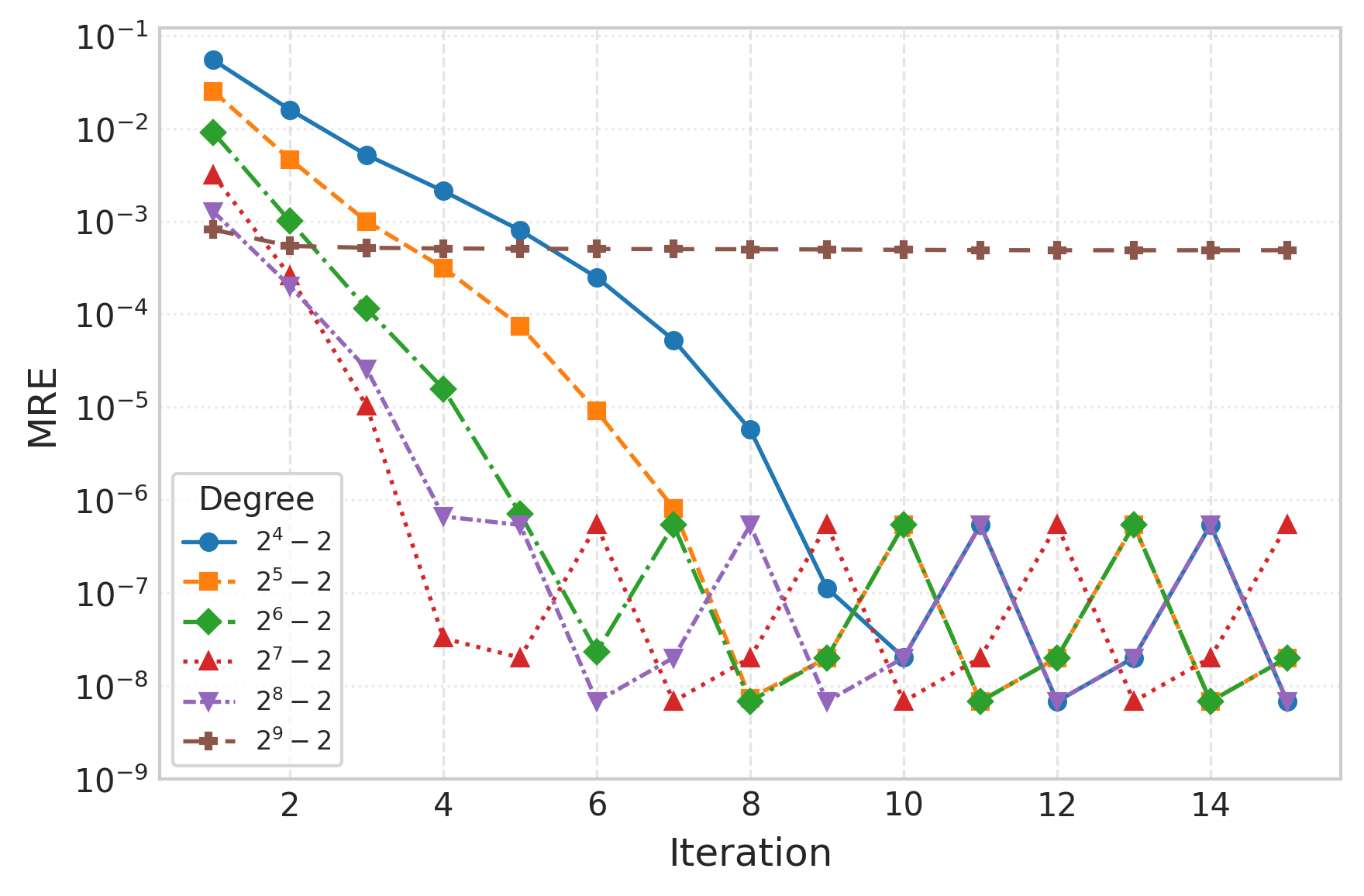}
    \Description{Visualization of Level 7 Case N}
    \caption{No Pre-BTS ($c=0$)}
    \label{fig:sub_a}
  \end{subfigure}
  \begin{subfigure}{0.49\linewidth}
    \centering
    \includegraphics[width=0.95\linewidth]{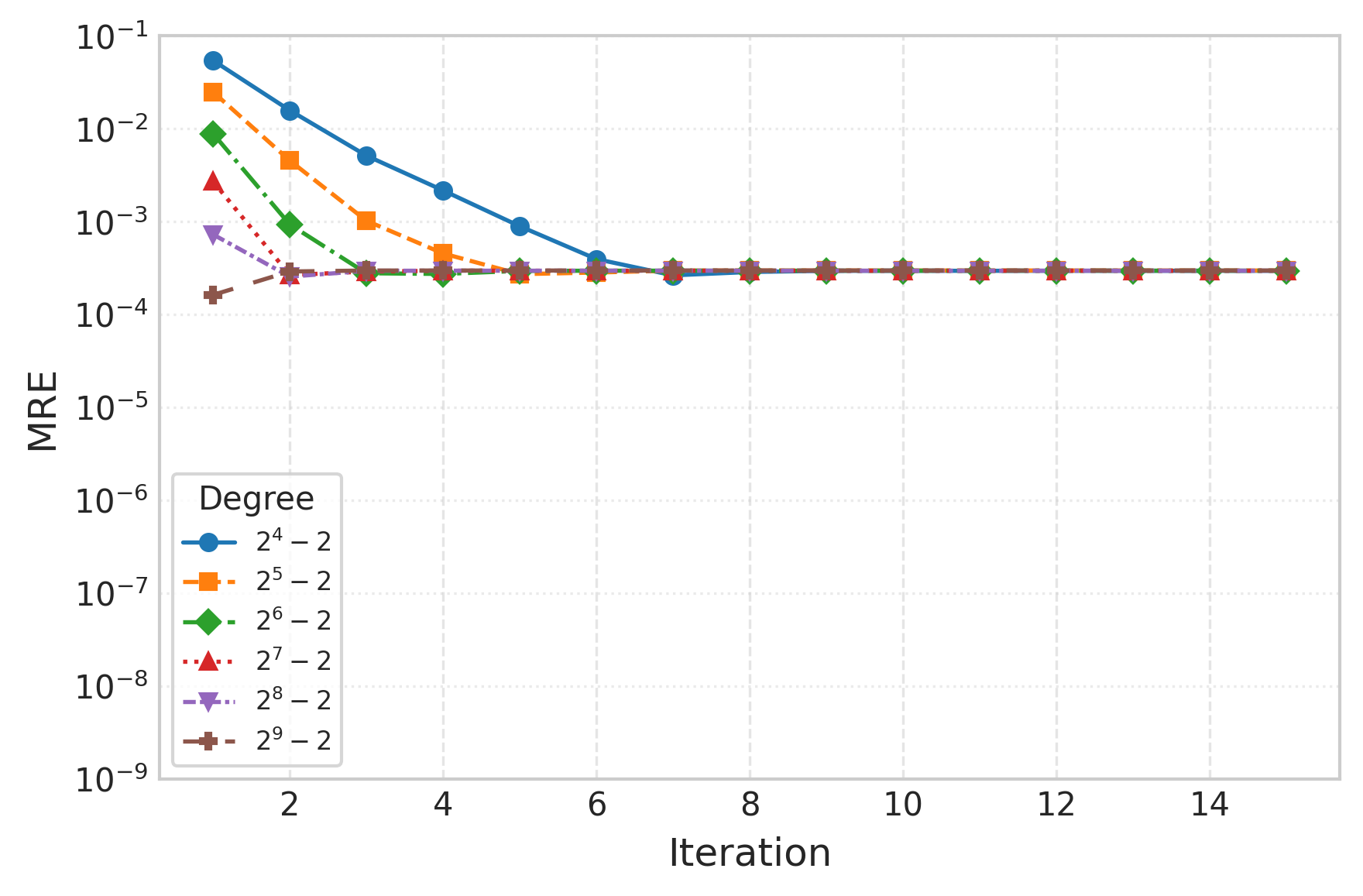}
    \Description{Visualization of Level 7 Case Y}
    \caption{Pre-BTS ($c=1$)}
    \label{fig:sub_b}
  \end{subfigure}

  \caption{MRE Curves for Newton's Method with Input Ciphertext Level $7$.}
  \label{fig:lattigo_level7}
\end{figure*}

Figure~\ref{fig:lattigo_level7} shows the MRE curves over 15 Newton iterations with the input ciphertext level set to $7$. Figure~\ref {fig:sub_a} corresponds to the case $c=0$, where $BTS$ is not applied before executing \textit{CryptoInvSqrt}. As observed, the MRE repeatedly exhibits a drastic increase followed by a decrease, except for degree $2^9-2$. The reason is that $BTS$ is applied during the iterations, and the error introduced by $BTS$ exceeds the error from Newton's method, resulting in a significant increase in the overall error. On the other hand, when the degree is $2^9-2$, the MRE is significantly higher than other degrees, around $10^{-4}$. In the evaluation of a Chebyshev polynomial of degree $2^{d_e}-2$, the computation consumes $d_e$ levels. When $d_e=9$, a $BTS$ must first be performed for the Chebyshev evaluation for a ciphertext at level 7. The $BTS$ error becomes increasingly significant as the degree grows, causing the Chebyshev evaluation error to rise significantly.

Figure~\ref {fig:sub_b} corresponds to the case $c=1$, where $BTS$ is applied before executing \textit{CryptoInvSqrt}. 
When applying $BTS$ to the input ciphertext $ct$, it produces a ciphertext $ct'$ that includes the $BTS$ error.
Executing \textit{CryptoInvSqrt} on $ct'$ then results in lower accuracy compared to performing the inverse square root directly on $ct$.
As observed in Figure ~\ref{fig:sub_a}, the minimum MRE for most degrees remains in the range of $[10^{-8}, 10^{-9}]$, whereas in Figure ~\ref{fig:sub_b}, it stays in the range of $[10^{-3}, 10^{-4}]$. However, as the ciphertext level decreases, the case $c=1$ shows a growing speed advantage over $c=0$, as confirmed in Table~\ref {table:lattigo_result}.

Table~\ref{table:lattigo_result} summarizes the identified optimal parameters of \textit{CryptoInvSqrt} for each possible ciphertext level using \sysname. Due to space limitations, only selected levels are presented.
In Algorithm ~\ref{alg:optimizing_params}, both $\theta$ and $\delta$ are set to 1, and scaling constant $B$ is set to 100.
For levels 11, 10, 2, and 1, the MRE contraint did not affect the choice of parameters. This is because levels 11 and 10 are evaluated only with $c=0$, and levels 2 and 1 are evaluated only with $c=1$.
As observed for level 7, when accuracy is prioritized (Y for MRE constraint), the optimal Pre-BTS indicator is $c=0$, whereas when speed is prioritized (N for MRE constraint), the optimal Pre-BTS indicator is $c=1$. We confirmed that \sysname performs parameter tuning and assigns different parameters to each ciphertext level on the target platform, while PP-STAT employed fixed parameters.

\begin{table}[!ht]
\caption{Optimization results of \sysname for all possible ciphertext in \lattigo. $c$ denotes the Pre-BTS indicator}
\centering
\footnotesize
\renewcommand{\arraystretch}{0.95}
\setlength{\tabcolsep}{4pt}
\begin{tabular}{ccccccc}
\toprule
\textbf{level} & \textbf{MRE constraint$^\ast$} & \textbf{degree} & $c$ & \textbf{iter} & \textbf{Time(s)} & \textbf{MRE} \\
\midrule
11  & Y & $2^7-2$ & 0 & 5 & 49.82 & $6.81 \cdot 10^{-9}$ \\
\midrule
10  & Y & $2^7-2$ & 0 & 5 & 49.12 & $2.51 \cdot 10^{-9}$ \\
\midrule
\multicolumn{7}{c}{$\cdots$} \\
\midrule
7  & Y & $2^6-2$ & 0 & 8 & 136.14 & $6.81 \cdot 10^{-9}$ \\
   & N & $2^5-2$ & 1 & 5 & 90.30  & $2.73 \cdot 10^{-4}$ \\
\midrule
\multicolumn{7}{c}{$\cdots$} \\
\midrule
2  & N & $2^6-2$ & 1 & 3 & 91.08  & $2.78 \cdot 10^{-4}$ \\
\midrule
1  & N & $2^6-2$ & 1 & 3 & 91.05  & $2.79 \cdot 10^{-4}$ \\
\bottomrule
\end{tabular}
\begin{flushleft}
$^\ast$ Y indicates that the bounded-MRE constraint is applied and $u_1$ is chosen from the output $R[l]$ of Algorithm~\ref{alg:optimizing_params}.
N indicates that $u_2$ is chosen.
\end{flushleft}
\label{table:lattigo_result}
\end{table}

%% file: section/6_Experiments.tex
\section{Performance Analysis}
\label{sec:experiments}

\subsection{Experimental Setup}
\label{subsec:experimental_setup}
We evaluate the impact of \sysname on precision and runtime under the CKKS setting. All experiments are conducted using HEaaN~\cite{heaan} for both CPU and GPU environments and Lattigo~\cite{lattigo}. \invSqrt and all statistical operations used to evaluate \sysname are implemented using the statistical computation modules of a representative HE-based framework.

\noindent\textbf{Hardware Configuration.}
Experiments on both \heaanCpu and \heaanGpu are conducted on a server with an Intel Core i9-10900X CPU, an NVIDIA GeForce RTX 3090 GPU, and 192GB RAM. Experiments on Lattigo are conducted on a server with an Intel Xeon Gold 6248R CPU and 64GB RAM.

\noindent\textbf{Parameter Settings.}
HEaaN supports various CKKS parameter configurations. In our experiments, we adopt the representative FGb setting, which provides bootstrapping support. The detailed parameters of HEaaN and Lattigo are described in Section~\ref{subsec:he_background}.
Throughout this section, we define $l$, $d$, $c$, and $i$ as the level, degree of Chebyshev polynomial, Pre-BTS indicator, and the number of Newton's iterations, respectively.



\subsection{Performance of Inverse Square Root} 
\label{subsec:InvSqrt}
We evaluate how much \sysname improves the accuracy and efficiency of \ppstat~\cite{choi2025pp} by comparing it with the existing SotA methods, \hstat~\cite{lee2023heaan} and PP-STAT.
\footnote{Note that \hstat in our experiments does not refer to the HEaaN.Stat library~\cite{heaanDocker} provided by CryptoLab, but rather denotes the implementation of the algorithm proposed in ~\cite{lee2023heaan}. For fair comparison, we re-implemented and executed both \hstat and \ppstat under the same experimental environment.}
In HEaaN-STAT, the number of iterations is fixed to $21$, and the original method supports only the domain $(0,1]$. For a fair comparison, we scale the inputs for HEaaN-STAT by a factor of $1/B$, and scale back the outputs by $\sqrt{B}$.
In PP-STAT, the degree of Chebyshev polynomials and the number of iterations are fixed at $2^9-2$, and 6 respectively.
We evaluated the MRE over $N/2=32768$ real values in the domain $[0.001, 100]$ and set the scaling constant $B$ to 100. Due to space limitations, we report only the results for ciphertext levels $l_{\max}$ and 7. At ciphertext level $l_{\max}$, Algotirhm ~\ref{alg:optimizing_params} outputs a single parameter combination. Therefore, we evaluated only this combination at level $l_{\max}$. All values in the tables are averaged over ten trials.

Table~\ref{table:comparison_invsqrt_hn_cpu} compares HEaaN-STAT, PP-STAT, and PP-STAT with \sysname on \heaanCpu. 
PP-STAT w/ \sysname(A) and PP-STAT w/ \sysname(S) represent the results of PP-STAT with accuracy-oriented parameters and speed-oriented parameters provided by \sysname, respectively.
At ciphertext level $12$, \sysname accelerated PP-STAT by $2.18\times$ with comparable MRE. Consequently, PP-STAT with \sysname achieved $2.21\times$ lower MRE than HEaaN-STAT, with $1.46\times$ faster runtime. 
At ciphertext level $7$, with accuracy-oriented parameters, 
\sysname accelerated PP-STAT by $1.99\times$ with comparable MRE.
PP-STAT with \sysname achieved $3.84\times$ lower MRE over HEaaN-STAT with $2.89\times$ faster runtime.
With speed-oriented parameters, PP-STAT with \sysname achieved $5.77\times$ faster runtime over HEaaN-STAT at the cost of $7.67\times$ increase in MRE, and $3.98\times$ faster runtime over the original PP-STAT at the cost of $29.48\times$ increase in MRE.

\begin{table}[h]
\scriptsize
\centering
\caption{Inverse square root computation on \heaanCpu.}
\renewcommand{\arraystretch}{0.95}
\setlength{\tabcolsep}{6pt}
\begin{tabular}{lcccccc}
\toprule
\textbf{Method} & \textbf{$l$} & \textbf{$d$} & \textbf{$c$} & \textbf{$i$} & \textbf{MRE} & \textbf{Time(s)} \\
\midrule
HEaaN-STAT~\cite{lee2023heaan}      & 12 & -        & - & 21 & $6.54 \times 10^{-8}$ & 26.93  \\
PP-STAT~\cite{choi2025pp}           & 12 & $2^9-2$  & - & 6  & $2.94 \times 10^{-8}$ & 40.23  \\
PP-STAT w/ \sysname                 & 12 & $2^5-2$  & 0 & 8  & $2.96 \times 10^{-8}$ & 18.47  \\
\midrule
HEaaN-STAT~\cite{lee2023heaan}      & 7  & -        & - & 21 & $1.18 \times 10^{-7}$ & 79.70  \\
PP-STAT~\cite{choi2025pp}           & 7  & $2^9-2$  & - & 6  & $3.07 \times 10^{-8}$ & 55.02  \\
PP-STAT w/ \sysname(A)              & 7  & $2^7-2$  & 0 & 5  & $3.07 \times 10^{-8}$ & 27.60  \\
PP-STAT w/ \sysname(S)              & 7  & $2^4-2$  & 1 & 10 & $9.05 \times 10^{-7}$ & 13.82  \\
\bottomrule
\end{tabular}
\label{table:comparison_invsqrt_hn_cpu}
\end{table}

Table~\ref{table:comparison_invsqrt_hn_gpu} compares HEaaN-STAT, PP-STAT, and PP-STAT with \sysname on \heaanGpu.
At ciphertext level $12$, \sysname accelerated PP-STAT by $2.35\times$ with comparable MRE. 
Consequently, PP-STAT with \sysname achieved $2.29\times$ lower MRE over HEaaN-STAT with $2.15\times$ faster runtime.
At ciphertext level $7$, with accuracy-oriented parameters, 
\sysname accelerated PP-STAT by $1.79\times$ with comparable MRE.
PP-STAT with \sysname achieved $3.76\times$ lower MRE over HEaaN-STAT with $3.26\times$ faster runtime.
With speed-oriented parameters, PP-STAT with \sysname achieved $6.34\times$ faster runtime over HEaaN-STAT at the cost of $8.04\times$ increase in MRE, and $3.49\times$ faster runtime over PP-STAT at the cost of $30.13\times$ increase in MRE.

\begin{table}[h]
\scriptsize
\centering
\caption{Inverse square root computation on \heaanGpu.}
\renewcommand{\arraystretch}{0.95}
\setlength{\tabcolsep}{6pt}
\begin{tabular}{lcccccc}
\toprule
\textbf{Method} & \textbf{$l$} & \textbf{$d$} & \textbf{$c$} & \textit{$i$} & \textbf{MRE} & \textbf{Time(s)} \\
\midrule
HEaaN-STAT~\cite{lee2023heaan}      & 12 & -        & - & 21 & $6.86 \times 10^{-8}$ & 0.73 \\
PP-STAT~\cite{choi2025pp}           & 12 & $2^9-2$  & - & 6  & $2.99 \times 10^{-8}$ & 0.80 \\
PP-STAT w/ \sysname                 & 12 & $2^8-2$  & 0 & 4  & $2.99 \times 10^{-8}$ & 0.34 \\
\midrule
HEaaN-STAT~\cite{lee2023heaan}      & 7  & -        & - & 21 & $1.15 \times 10^{-7}$ & 2.22  \\
PP-STAT~\cite{choi2025pp}           & 7  & $2^9-2$  & - & 6  & $3.07 \times 10^{-8}$ & 1.22  \\
PP-STAT w/ \sysname(A)              & 7  & $2^7-2$  & 0 & 5  & $3.06 \times 10^{-8}$ & 0.68 \\
PP-STAT w/ \sysname(S)              & 7  & $2^7-2$  & 1 & 4  & $9.25 \times 10^{-7}$ & 0.35 \\
\bottomrule
\end{tabular}
\label{table:comparison_invsqrt_hn_gpu}
\end{table}

Notably, Table ~\ref{table:comparison_invsqrt_lattigo} demonstrates that \sysname substantially improves both speed and accuracy on \lattigo.
At ciphertext level $11$, \sysname accelerated PP-STAT by $1.96\times$ while simultaneously reducing the MRE by $79.00 \times$. Consequently, PP-STAT with \sysname achieved $7.75 \cdot 10^{5} \times$ lower MRE over HEaaN-STAT, with $4.63\times$ faster runtime.
At ciphertext level $7$, with accuracy-oriented parameters, PP-STAT with \sysname achieved an MRE that was $7.80 \cdot 10^{5} \times$ lower than HEaaN-STAT with $3.32\times$ faster runtime, and $7.44 \cdot 10^{4} \times$ lower MRE than PP-STAT with $1.05\times$ faster runtime. 
With speed-oriented parameters, PP-STAT with \sysname achieved 
$4.91\times$ and $1.54\times$ faster runtimes over HEaaN-STAT and PP-STAT, respectively.
Notably, it also reduced MRE significantly
($19.45\times$ and $1.86\times$).


\begin{table}[h]
\scriptsize
\centering
\caption{Inverse square root computation on \lattigo.}
\renewcommand{\arraystretch}{0.95}
\setlength{\tabcolsep}{6pt}
\begin{tabular}{lcccccc}
\toprule
\textbf{Method} & \textbf{$l$} & \textbf{$d$} & \textbf{$c$} & \textbf{$i$} & \textbf{MRE} & \textbf{Time(s)} \\
\midrule
HEaaN-STAT~\cite{lee2023heaan}      & 11 & -        & - & 21 & $5.28 \times 10^{-3}$ & 228.82  \\
PP-STAT~\cite{choi2025pp}           & 11 & $2^9-2$  & - & 6  & $5.38 \times 10^{-7}$ & 96.67 \\
PP-STAT w/ \sysname                 & 11 & $2^7-2$  & 0 & 5  & $6.81 \times 10^{-9}$ & 49.42 \\
\midrule
HEaaN-STAT~\cite{lee2023heaan}      & 7  & -        & - & 21 & $5.31 \times 10^{-3}$ & 448.30  \\
PP-STAT~\cite{choi2025pp}           & 7  & $2^9-2$  & - & 6  & $5.07 \times 10^{-4}$ & 141.04  \\
PP-STAT w/ \sysname (A)             & 7  & $2^6-2$  & 0 & 8  & $6.81 \times 10^{-9}$ & 134.87  \\
PP-STAT w/ \sysname (S)             & 7  & $2^5-2$  & 1 & 5  & $2.73 \times 10^{-4}$ & 91.34 \\
\bottomrule
\end{tabular}
\label{table:comparison_invsqrt_lattigo}
\end{table}


\subsection{Performance of Statistical Operations} 
\label{subsec:STAT}

\begin{table}[!ht]
\footnotesize
\centering
\caption{Performance of statistical measures on \heaanCpu.}
\setlength{\tabcolsep}{4pt}
\begin{tabular}{lccccc}
\toprule
\textbf{Measure} & \textbf{Method} & \textbf{$c$} & \textbf{$B$} & \textbf{MRE} & \textbf{Time (s)} \\
\midrule
\multirow{3}{*}{ZNorm} & PP-STAT~\cite{choi2025pp}  & - & 100 & $8.91 \times 10^{-6}$ & 89.43 \\
                       & PP-STAT w/ \sysname(A)     & 0 & 100 & $8.91 \times 10^{-6}$ & 68.32 \\
                       & PP-STAT w/ \sysname(S)     & 1 & 100 & $1.19 \times 10^{-5}$ & 38.65 \\
\midrule

\multirow{3}{*}{Skew}  & PP-STAT~\cite{choi2025pp}  & - & 20  & $2.46 \times 10^{-4}$ & 102.96 \\
                       & PP-STAT w/ \sysname(A)     & 0 & 20  & $2.46 \times 10^{-4}$ & 76.29 \\
                       & PP-STAT w/ \sysname(S)     & 1 & 20  & $2.39 \times 10^{-4}$ & 48.57 \\
\midrule

\multirow{3}{*}{Kurt}  & PP-STAT~\cite{choi2025pp}  & - & 20  & $5.05 \times 10^{-5}$ & 102.90 \\
                       & PP-STAT w/ \sysname(A)     & 0 & 20  & $3.89 \times 10^{-5}$ & 75.50 \\
                       & PP-STAT w/ \sysname(S)     & 1 & 20  & $7.33 \times 10^{-3}$ & 45.83 \\
\midrule

\multirow{3}{*}{PCC}   & PP-STAT~\cite{choi2025pp}  & - &
20  & $1.33 \times 10^{-6}$ & 198.40 \\
                       & PP-STAT w/ \sysname(A)     & 0 & 20  & $1.37 \times 10^{-6}$ & 143.53 \\
                       & PP-STAT w/ \sysname(S)     & 1 & 20  & $5.95 \times 10^{-6}$ & 84.32  \\
\bottomrule
\end{tabular}
\label{table:stat_result_hn_cpu}
\end{table}

We evaluate \sysname on the statistical operations in PP-STAT to verify the effect of parameter optimization. These operations include Z-score normalization (ZNorm), skewness (Skew), kurtosis (Kurt), and the Pearson correlation coefficient (PCC).
They involve computing the inverse square root of the variance. We randomly sampled 1,000,000 real values so that the variance of the sampled dataset lies within $[0.001, B]$. 
ZNorm is evaluated over $[0, 100]$, and the other measures over $[0, 20]$. In this section, only the experimental results for \heaanCpu and \heaanGpu are presented, while the results for \lattigo are omitted due to space limitations.
Instead, Section ~\ref{subsec:real_dataset} will provide the experimental results on \lattigo with real-world datasets, demonstrating the efficiency of \sysname on Lattigo.
Only the results for ciphertext level $7$ are shown due to limited space. All values in the tables are averaged over ten trials.

Table~\ref{table:stat_result_hn_cpu} compares PP-STAT with and without \sysname on \heaanCpu.
First, when comparing PP-STAT with accuracy-oriented \sysname(A), both methods yield nearly identical MREs, while PP-STAT with \sysname(A) runs $1.31\times$ faster for ZNorm, $1.35\times$ for Skew, $1.36\times$ for Kurt, and $1.38\times$ for PCC. Next, when comparing PP-STAT with speed-oriented \sysname(S), it runs $2.31\times$ faster for ZNorm, $2.12\times$ for Skew, $2.25\times$ for Kurt, and $2.35\times$ for PCC, while maintaining comparable MREs for ZNorm and Skew, with higher MREs observed for Kurt ($145\times$) and PCC ($4.47\times$).

Table~\ref{table:stat_result_hn_gpu} summarizes the results on \heaanGpu. When comparing PP-STAT with \sysname(A), the two methods show nearly identical MREs, while PP-STAT with \sysname(A) runs $1.14\times$ faster for ZNorm, $1.24\times$ for Skew, $1.24\times$ for Kurt, and $1.25\times$ for PCC. Next, when comparing PP-STAT with \sysname(S), it runs $2.12\times$ faster for ZNorm, $1.98\times$ for Skew, $1.97\times$ for Kurt, and $2.05\times$ for PCC, while maintaining comparable MREs for ZNorm and Skew, with higher MREs observed for Kurt ($2.48\times$) and PCC ($15.68\times$).

\begin{table}[!ht]
\footnotesize
\centering
\caption{Performance of statistical measures on \heaanGpu.}
\begin{tabular}{lccccc}
\toprule
\textbf{Measure} & \textbf{Method} & \textbf{$c$} & \textbf{$B$} & \textbf{MRE} & \textbf{Time (s)} \\
\midrule
\multirow{3}{*}{ZNorm} & PP-STAT~\cite{choi2025pp}  & - & 100 & $3.76 \times 10^{-6}$ & 2.63 \\
                       & PP-STAT w/ \sysname(A)     & 0 & 100 & $3.76 \times 10^{-6}$ & 2.30 \\
                       & PP-STAT w/ \sysname(S)     & 1 & 100 & $4.72 \times 10^{-6}$ & 1.24 \\
\midrule

\multirow{3}{*}{Skew}  & PP-STAT~\cite{choi2025pp}  & - & 20  & $2.83 \times 10^{-4}$ & 2.85 \\
                       & PP-STAT w/ \sysname(A)     & 0 & 20  & $2.83 \times 10^{-4}$ & 2.30 \\
                       & PP-STAT w/ \sysname(S)     & 1 & 20  & $2.73 \times 10^{-4}$ & 1.44 \\
\midrule

\multirow{3}{*}{Kurt}  & PP-STAT~\cite{choi2025pp}  & - & 20  & $2.54 \times 10^{-3}$ & 2.83 \\
                       & PP-STAT w/ \sysname(A)     & 0 & 20  & $2.55 \times 10^{-3}$ & 2.29 \\
                       & PP-STAT \sysname(S)        & 1 & 20  & $6.30 \times 10^{-3}$ & 1.44 \\
\midrule

\multirow{3}{*}{PCC}   & PP-STAT~\cite{choi2025pp}  & - & 20  & $3.03 \times 10^{-6}$ & 5.48 \\
                       & PP-STAT w/ \sysname(A)     & 0 & 20  & $3.01 \times 10^{-6}$ & 4.38 \\
                       & PP-STAT w/ \sysname(S)     & 1 & 20  & $4.75 \times 10^{-5}$ & 2.67 \\
\bottomrule
\end{tabular}
\label{table:stat_result_hn_gpu}
\end{table}

\subsection{Evaluation on Real-World Datasets}
\label{subsec:real_dataset}
We evaluate the performance of \sysname using four statistical measures---ZNorm, Skew, Kurt, and PCC---on the same real-world datasets used in prior work. Two widely used datasets are employed: the UCI Adult Income dataset~\cite{adult_2} (\textit{Adult}) and the Medical Cost Insurance dataset~\cite{akter2018investigation} (\textit{Insurance}). All experiments are conducted on \lattigo. Due to space constraints, results are reported only for ciphertext level 11, where Algorithm~\ref{alg:optimizing_params} outputs a single parameter combination. All values in the tables are averaged over ten trials.

\noindent\textbf{Datasets.}
The \textit{Adult} dataset includes 48,842 records with 14 features, such as \texttt{age}, \texttt{hours-per-week}, and \texttt{education-num}. The \textit{Insurance} dataset contains 1,338 records with 7 attributes, including \texttt{age}, \texttt{bmi}, \texttt{smoker}, and \texttt{charges}.

\noindent\textbf{Results on \textit{Adult}.}
We select three continuous-valued features—\texttt{age}, \texttt{hours-per-week}, and \texttt{education-num}. We compute ZNorm, Skew, and Kurt for each feature. 
PCC is measured between \texttt{age} and the other two features. Table~\ref{table:adult_evaluation} summarizes the results, showing that while the MREs of both systems remain similar, PP-STAT with \sysname achieves $42\%$--$44\%$ faster execution time than the baseline.

\begin{table}[!ht]
\centering
\caption{Performance of statistical measures over the \textit{Adult} dataset (with fixed scaling factor $B=50$). Runtime reduction (R) is computed as $(1 - (b) / (a)) \times 100\%$. Kurtosis is reported as excess kurtosis (i.e., normal kurtosis minus 3).}
\label{table:adult_evaluation}
\resizebox{1\linewidth}{!}{
\begin{tabular}{llccccc}
\toprule
\multirow{2}{*}{\textbf{Measure}} & \multirow{2}{*}{\textbf{Feature(s)}} &
\multicolumn{2}{c}{\textbf{PP-STAT~\cite{choi2025pp}}} &
\multicolumn{3}{c}{{\textbf{PP-STAT w/ \sysname}}} \\
\cmidrule(lr){3-4} \cmidrule(lr){5-7}
 &  & \textbf{MRE} & $(a)$\textbf{Runtime (s)} & \textbf{MRE} & $(b)$\textbf{Runtime (s)} & \textbf{R(\%)} \\
\midrule

\multirow{3}{*}{ZNorm} 
  & AGE & $2.82 \times 10^{-8}$ & 110.83 & $2.12 \times 10^{-8}$ & 62.89 & 44.12 \\
  & EDU & $5.21 \times 10^{-8}$ & 109.86 & $5.30 \times 10^{-8}$ & 61.02 & 44.00 \\
  & HPW & $5.93 \times 10^{-8}$ & 109.47 & $5.23 \times 10^{-8}$ & 61.09 & 44.16 \\
\midrule

\multirow{3}{*}{Skew}
  & AGE & $5.97 \times 10^{-8}$ & 112.86 & $5.92 \times 10^{-8}$ & 64.25 & 42.47 \\
  & EDU & $8.63 \times 10^{-8}$ & 112.91 & $8.88 \times 10^{-8}$ & 63.49 & 42.70 \\
  & HPW & $1.04 \times 10^{-7}$ & 112.43 & $4.63 \times 10^{-8}$ & 63.48 & 43.63 \\
\midrule

\multirow{3}{*}{Kurt}
  & AGE & $5.21 \times 10^{-6}$ & 113.00 & $1.20 \times 10^{-6}$ & 63.75 & 44.27 \\
  & EDU & $6.41 \times 10^{-7}$ & 112.87 & $6.41 \times 10^{-7}$ & 64.16 & 42.83 \\
  & HPW & $3.54 \times 10^{-7}$ & 112.79 & $6.41 \times 10^{-8}$ & 63.09 & 43.77 \\
\midrule

\multirow{2}{*}{PCC} 
  & AGE vs HPW & $2.50 \times 10^{-8}$ & 223.32 & $3.39 \times 10^{-8}$ & 124.86 & 43.43 \\
  & AGE vs EDU & $2.65 \times 10^{-8}$ & 223.41 & $4.65 \times 10^{-8}$ & 124.49 & 43.96 \\
\bottomrule
\end{tabular}
}
\end{table}

\noindent\textbf{Results on \textit{Insurance}}
Following PP-STAT, we select three attributes: \texttt{age}, \texttt{bmi}, and \texttt{smoker}. Other categorical or discrete fields are omitted in order to simplify encrypted processing. The variables \texttt{age} and \texttt{bmi} are continuous, while the binary feature \texttt{smoker} is encoded numerically (\texttt{yes}=1, \texttt{no}=0). The target variable \texttt{charges} ranges from $1121.87$ to $63770.43$. To align with the CKKS scale, this feature is divided by 1000 before encryption. Since PCC is scale-invariant, scaling does not affect the PCC results.
In Table~\ref {table:insurance_evaluation}, ZNorm, Skew, and Kurt are evaluated only on the \texttt{charges}, while PCC is evaluated between \texttt{charges} and the other three features. 
As shown in the table, PP-STAT with \sysname achieves up to $44.6\%$ faster runtime than PP-STAT,
while the MREs 
remain similar. 
\begin{table}[!ht]
\centering
\caption{Performance of statistical measures over the \textit{Insurance} dataset. The scaling factor $B$ is set to $100$ for Z-score normalization and $20$ for all other evaluations. Runtime reduction (R) is computed as $(1 - (b) / (a)) \times 100\%$. Kurtosis is reported as excess kurtosis (i.e., normal kurtosis minus 3).}
\label{table:insurance_evaluation}
\resizebox{1\linewidth}{!}{
\begin{tabular}{llccccc}
\toprule
\multirow{2}{*}{\textbf{Measure}} & \multirow{2}{*}{\textbf{Feature(s)}} &
\multicolumn{2}{c}{\textbf{PP-STAT~\cite{choi2025pp}}} &
\multicolumn{3}{c}{{\textbf{PP-STAT w/ \sysname}}} \\
\cmidrule(lr){3-4} \cmidrule(lr){5-7}
 &  & \textbf{MRE} & $(a)$\textbf{Runtime (s)} & \textbf{MRE} & $(b)$\textbf{Runtime (s)} & \textbf{R(\%)} \\
\midrule

ZNorm 
  & Charges & $1.88 \times 10^{-8}$ & 110.29 & $1.22 \times 10^{-8}$ & 61.07 & 44.64 \\
\midrule

Skew
  & Charges & $5.92 \times 10^{-8}$ & 112.34 & $3.74 \times 10^{-8}$ & 63.41 & 43.56 \\
\midrule

Kurt
  & Charges & $2.96 \times 10^{-7}$ & 112.32 & $1.07 \times 10^{-7}$ & 63.56 & 43.43 \\
\midrule

\multirow{3}{*}{PCC} 
  & AGE vs Charges    & $1.84 \times 10^{-8}$ & 222.26 & $2.97 \times 10^{-8}$ & 124.49 & 43.71 \\
  & BMI vs Charges    & $3.29 \times 10^{-8}$ & 222.64 & $4.01 \times 10^{-8}$ & 124.15 & 44.22 \\
  & Smoker vs Charges & $1.22 \times 10^{-8}$ & 222.74 & $2.89 \times 10^{-8}$ & 123.97 & 44.31 \\
\bottomrule
\end{tabular}
}
\footnotesize{\textit{Abbreviations:} AGE = age, BMI = body mass index}
\end{table}

%% file: section/7_Discussion.tex
\section{Discussion and Future Work}
\label{sec:discussion_future_work}
The current design determines the optimal parameters by evaluating all possible scenarios on a given platform.
However, the exhaustive evaluation is highly time-consuming in CPU-based libraries. Table~\ref{table:init_time} presents the exploration time of \sysname across three libraries: \heaanCpu, \heaanGpu, and \lattigo. As we used different parameter settings for each library, we conducted exhaustive exploration for 11 levels in \lattigo and 9 levels in \heaan.

\begin{table}[h]
\footnotesize
\centering
\caption{Exploration time of \sysname.}
\renewcommand{\arraystretch}{0.95}
\setlength{\tabcolsep}{6pt}
\begin{tabular}{lccc}
\toprule
\textbf{Libraries} & \heaanCpu & \heaanGpu & \lattigo \\
\midrule
Runtime (s)   & 3178.89 & 117.33 & 30923.34 \\
\bottomrule
\end{tabular}
\label{table:init_time}
\end{table}

In Table~\ref{table:init_time}, the exploration time of \sysname is relatively high in \lattigo.
The reason is that \sysname repeatedly invokes bootstrapping and Chebyshev polynomial evaluation, which are the most computationally expensive operations in our pipeline, especially in \lattigo. \sysname requires re-tuning the parameters whenever the underlying hardware changes (e.g., CPU–GPU switching) or when the library version is updated.
The considerable execution time of the parameter selection can negatively impact the overall system performance, especially during re-tuning.

Thus, in future work, we plan to develop a strategy to reduce the total exploration time by identifying the optimal parameters using easily measurable information, such as computational errors, instead of relying on full scenario evaluations.

In this paper, we did not include the coefficient of variation (CV), although it was considered in PP-STAT~\cite{choi2025pp}. 
In CV, the denominator may become negative.
Thus, a sign function (outputting 1 for non-negative inputs and -1 otherwise) is applied before using \invSqrt to ensure non-negativity. Because the performance depends on the specific implementation of the sign function, it is considered out of scope and excluded from this study.



%% file: section/3_Related_work.tex
\section{Related Work}
\label{sec:relatedwork}
Lee et al.~\cite{lee2023heaan} introduced HEaaN-STAT, an HE-based statistical framework. It employs Newton’s method for the inverse square root operation with a constant initial value and shows that the approximation error converges quadratically to zero. Their experiments were conducted on a GPU environment using a custom-developed library.

Panda~\cite{panda2022polynomial} proposed the Pivot–Tangent method, which selects a suitable initial value for Newton’s method by adopting a sign function. They implemented their approach in SEAL~\cite{sealcrypto}. Since SEAL does not support bootstrapping, they manually decrypted and re-encrypted the ciphertext once it reached level~0, conducting all experiments on a CPU environment.

Qu and Xu~\cite{qu2023improvements} argued that minimax polynomials are unsuitable for selecting initial values in Newton’s method. Instead, they proposed rational and Taylor expansion-based initialization strategies for the inverse square root, with experiments conducted in SEAL on a CPU environment.

Recently, Choi~\cite{choi2025pp} proposed PP-STAT, a privacy-preserving statistical analytics framework under HE. It employed Chebyshev approximation to determine an appropriate initial value for Newton’s method and introduced a pre-normalization scaling technique to reduce overall level consumption. All experiments were performed on Lattigo~\cite{lattigo} in a CPU environment.

In summary, these studies have explored various optimizations of Newton’s method for inverse square root computation under HE. However, all of them restricted their evaluations to a single library and environment with fixed HE parameter settings.

%% file: section/Security_Analysis.tex
\section{Security Analysis}
\label{sec:security_analysis}

We demonstrate that \sysname protects the client’s data against a semi-honest (honest-but-curious) server~\cite{chandran2022simc}, which is one of the standard adversarial models in HE-based service architectures~\cite{choi2024blindtouch, choi2024blindmatch, lee2023hetal}. During the provisioning phase, \sysname generates the public, secret, and evaluation keys and uses them to determine the optimized parameters for \invSqrt. Throughout this process, the server observes only encrypted data and does not have access to the client's plaintext or secret key. Therefore, \sysname protects the client’s data against a semi-honest server due to the semantic security of the CKKS scheme.
Even in the subsequent statistical analysis phase, all numerical operations (e.g., Chebyshev approximation, Newton’s iteration, and statistical measure evaluations such as ZNorm, Kurt, and PCC) are performed directly on ciphertexts. The server can observe only ciphertext structures and computational metadata (e.g., ciphertext level or slot size), which are independent of the underlying plaintext values and thus leak no meaningful information.
Although a malicious attacker could attempt to alter the parameter settings of \sysname,
we exclude this threat model involving a malicious server, because it is reasonable to assume that the server would not perform manipulations that compromise its own trustworthiness and business.
Consequently, \sysname ensures the confidentiality of both the client’s input data and intermediate results against a semi-honest server, inheriting the semantic security guarantees of the CKKS encryption scheme.

%% file: section/8_Conclusion.tex
\section{Conclusion}
\label{sec:conclusion}
We presented \sysname, an adaptive parameter optimizer that automatically selects optimal configurations for \textit{CryptoInvSqrt} across diverse HE platforms. By systematically evaluating possible parameter settings, \sysname achieves up to $2.35\times$ speedup on \heaanGpu, $2.18\times$ on \heaanCpu, and $1.96\times$ on \lattigo compared to PP-STAT~\cite{choi2025pp}, and up to $4.63\times$ faster execution than HEaaN-STAT~\cite{lee2023heaan}, while maintaining comparable accuracy.  
Applying the optimized parameters to four statistical operations (Z-score normalization, skewness, kurtosis, and Pearson correlation coefficient) confirms consistent acceleration and accuracy retention.  
Overall, \sysname provides a unified, platform-adaptive optimization framework that delivers high efficiency and precision for practical homomorphic encryption workloads.

%% file: section/Acknowledgments.tex
\section*{Acknowledgments}
We thank the anonymous reviewers for their valuable comments.
Hyunmin Choi and Mun-Kyu Lee are the corresponding authors.
This manuscript was entirely written by the authors. During its preparation, we used ChatGPT solely for grammar correction, word-choice refinement, and improvement of sentence clarity. This work was supported in part by the Institute of Information \& Communications Technology Planning \& Evaluation (IITP) grants funded by the Korea government (MSIT) [No.RS-2024-00436936 ($D^2$ (Disinformation \& Deepfake) Research Center)
and No.RS-2022-00155915 (Artificial Intelligence Convergence Innovation Human Resources Development (Inha University))]
and in part by an Inha University Research Grant (2024).